\documentstyle[12pt]{ioplppt}

\begin{document}
\jl{1}
\title{Directed Percolation with a Wall or Edge}
\author{Per Fr\"ojdh\dag,
        Martin Howard\ddag, 
        Kent B{\ae}kgaard Lauritsen\ddag}

\address{\dag NORDITA, Blegdamsvej 17, 2100 Copenhagen \O, Denmark}
\address{\ddag Center for Chaos and Turbulence Studies, The Niels Bohr 
         Institute, Blegdamsvej 17, 2100 Copenhagen \O, Denmark}

\begin{abstract}
We examine the effects of introducing a wall or edge into a directed 
percolation process. Scaling ansatzes are presented for the density and
survival probability of a cluster in these geometries, and we make the
connection to surface critical phenomena and field theory. The results
of previous numerical work for a wall can thus be interpreted in terms
of surface exponents satisfying scaling relations generalising those for
ordinary directed percolation.
New exponents for edge directed percolation are also introduced. They are
calculated in mean-field theory and measured numerically in $2+1$ dimensions. 
\end{abstract}

\maketitle

\section{Introduction}

The impact of boundaries on critical phenomena has been the focus of much 
research in recent years (for extensive reviews of
surface critical phenomena see \cite{binder,diehl}). 
In the presence of a boundary certain (surface) quantities no longer 
scale as in the bulk, but possess different exponents which are dependent on 
the boundary conditions. In the past, most research has focused on the 
effects of surfaces in equilibrium critical phenomena and far less 
attention has been paid to boundaries in dynamical systems, such 
as in directed percolation (DP). The DP universality class is thought
to describe a variety of phase transitions from non-trivial 
active into absorbing states \cite{kinzel} in processes such as epidemics,
chemical reactions \cite{grassberger-torre,schloegl}, catalysis \cite{zgb},
the contact process \cite{harris}, and certain cellular automata 
\cite{domany1,domany2}.
Since all of these physical systems contain boundaries, an understanding
of surface effects is very important.

The microscopic rules for bulk (bond) DP in $d+1$ dimensions are extremely 
simple: any site at time $t$ may make a connection to any of its 
$2d$ nearest neighbours at time $t+1$ with growth probability $p$. 
Below a threshold, $\Delta = p - p_c < 0$, such a process
always dies, whereas for $\Delta > 0$ there is a finite probability for
survival. At the transition point, the system is critical and scales 
anisotropically, i.e.\ the correlation lengths in time ($\parallel$) 
and space ($\perp$) scale with different exponents, $\xi_\parallel 
\sim |\Delta|^{-\nu_\parallel}$ and $\xi_\perp \sim |\Delta|^{-\nu_
\perp}$, respectively. 

Above the upper critical dimension, $d_c=4$, these exponents can be
calculated using a simple mean-field theory. However for $d<d_c$,
fluctuation effects become important, and hence the computation of
the exponents becomes a much harder task. The principal analytic
technique for this calculation employs the
equivalence between DP and Reggeon field theory \cite{cardy-sugar}. 
Using renormalisation group techniques, the exponents can then be 
computed perturbatively in an $\epsilon=d_c-d$ expansion. These 
analytic techniques are supplemented by simulations and series expansions, 
which mean that, for example, the bulk critical exponents for $d=1$
are known rather accurately \cite{iwan}. Nevertheless, 
an exact solution for DP remains an open, and extremely important, 
problem.

In this paper we will be exclusively interested in the effects of 
boundaries on DP clusters. In order to isolate their effects, it is 
convenient to consider a semi-infinite system, where the cluster 
grows from a seed close to the surface. Series expansions 
\cite{essam-etal:1996} and numerical simulations \cite{lauritsen-etal} 
in 1+1 dimensions indicate that the presence of the wall alters 
several exponents. In particular, the percolation probability (order
parameter),
\begin{equation}
P_1 (\Delta) \sim \Delta^{\beta_1} , \ \ \Delta \geq 0, 
\label{perc_prob} 
\end{equation}
scales with an exponent $\beta_1$ rather than the standard 
exponent $\beta$ (the subscript `1' refers to the wall).
However, the scaling properties of the correlation 
lengths (as given by $\nu_\parallel$ 
and $\nu_\perp$) are {\it not\/} altered. More surprising, however, is the 
appearance of an apparently integer exponent describing the mean lifetime
of a finite cluster in the presence of a wall:
\begin{equation}
\label{meantime}
\langle t \rangle \sim |\Delta|^{-\tau_1} . 
\end{equation}
Here $\tau_1 = \nu_\parallel - \beta_1 = 1.0002 \pm 0.0003$ in $1+1$ 
dimensions and is conjectured to be exactly unity \cite{essam-etal:1996}. 
If true, this would be a remarkable result, since none of the other 
exponents for DP are known exactly, and one even lacks evidence for them to 
be rational numbers. Note, however, that this situation is very different 
from the case of compact directed percolation, where no vacancies 
within a cluster are allowed. This model is relatively simple to solve 
and most of the exponents, including $\tau_1$, are integers (see 
\cite{essam-guttmann} and references therein).

The purpose of the present paper is to analyse the above results in the 
context of surface critical phenomena. We first of all write down
scaling ansatzes for the survival probability and cluster density which 
take boundary effects 
into account. From this we are able to derive the behaviour of,
for example, the cluster mass in terms of surface 
(and bulk) exponents. We emphasise that the new 
exponent $\beta_1$ describes the scaling of 
activity on the wall, thus $\beta_1$ is a so-called {\it surface\/} 
exponent \cite{binder,diehl}. We next consider 
the appropriate field theory for DP in a semi-infinite geometry. This
theory was first analysed by Janssen et al.\ \cite{janssen-etal}, where 
the appropriate surface exponents were computed to first order 
in $\epsilon = d_c-d$ using renormalisation group techniques. The 
introduction of an inactive wall results in the so-called ordinary 
surface transition at the bulk critical point, for which 
sites close to the wall are less likely 
to be active than those in the bulk. In this picture, it is clear that 
only certain exponents are altered: more precisely, the boundary 
introduces one new independent exponent, while all the bulk exponents 
remain unaffected. By placing a finite seed close to the boundary, 
however, the distribution functions of the emerging clusters become 
sensitive to the new exponent. Unfortunately, these field 
theoretic methods yield little insight into why $\tau_1$ should equal 
unity in $1+1$ dimensions. Hence we conclude that the apparently 
integer value for $\tau_1$ must be a special property of DP in $1+1$ 
dimensions, inaccessible to perturbative expansions about $d_c$. 

Finally, we extend our analysis by allowing the wall to have an edge
with opening angle $\alpha$. This leads to the introduction of new 
angle-dependent {\it edge\/} exponents which  
govern the properties of clusters started close to the edge.
We solve the corresponding mean-field theory and also determine the
exponents numerically in $2+1$ dimensions using computer simulations.

\section{Wall Analysis}

In this section we shall discuss the exponents associated with the growth 
of DP clusters in the presence of a wall. Some of our analysis will be 
similar in spirit to that of Grassberger \cite{grassberger-3d}, who 
analysed the case of ordinary percolation in a semi-infinite system. 

First of all, let us examine the effects of introducing a $d-1$ dimensional 
wall at $x_\perp=0$ [${\bf x}=({\bf x_{\parallel}},x_{\perp}=0)$] into a
DP process. Note that the labels parallel ($\parallel$) and perpendicular 
($\perp$) refer here to directions relative to the wall (and not 
relative to the time direction). Consider a cluster arising from a single 
seed located next to the wall at $t=0$. The probability that an infinite 
cluster can be grown from this seed is given by the percolation 
probability (\ref{perc_prob}) which scales as $\Delta^{\beta_1}$. 
Furthermore the probability that a surface point at a later time
belongs to this infinite cluster scales in the same way.
Thus $\beta_1$ is an (independent) surface exponent in analogy with 
surface critical phenomena for equilibrium statistical mechanics 
\cite{binder,diehl} (more details can be found in section 3, where we 
will also discuss how the surface scaling $\Delta^{\beta_1}$ crosses 
over to the bulk scaling $\Delta^{\beta}$).

For a given bulk universality class (such as that of DP), several 
surface universality classes are possible. In our case, the lattice has
simply been cut off and hence there will be fewer active points close to 
the surface. This corresponds to the boundary condition for the 
so-called ordinary transition (for which $\beta_1 > \beta$). 
The survival probability (the probability that the cluster is still alive 
at time $t$) has the form
\begin{equation}
	P_1(t,\Delta) = \Delta^{\beta_1} \, \psi_1 \left({{t}/
	{\xi_{\parallel}}}\right),            \label{eq:P_1(t,Delta)}
\end{equation}
where the scaling function $\psi_1$ is constant for $t \gg \xi_\parallel$
\cite{grassberger-torre}. Furthermore the presence of the wall leaves the 
scaling of the correlation lengths unaltered, and hence exponents such 
as $\nu_\perp$ and $\nu_\parallel$ are everywhere unaffected.
The mean lifetime of finite clusters (\ref{meantime}) follows from 
(\ref{eq:P_1(t,Delta)}) by averaging $t$ with respect to the cluster 
lifetime distribution $-d P_1/dt$. As a result,
\begin{equation}
	\label{tau1_relation}
	\tau_1 = \nu_\parallel - \beta_1.
\end{equation}
However, for $\nu_\parallel < \beta_1$, the leading contribution to 
$\langle t \rangle$ will be a constant, such that the above scaling relation 
breaks down and is replaced by $\tau_1 = 0$.
Note, however, if one instead considers a space-time geometry where 
the wall direction departs from the time direction, then all the above
quantities will, as usual, crossover to bulk scaling (see also 
\cite{essam-etal:1996}).

For the density $\rho_1$ of a cluster growing from a single seed located next 
to the wall we make the scaling ansatz
\begin{equation}
	\label{ansatz_rho_wall}
	\rho_{1}(x,t,\Delta) = \Delta^{\beta_1 + \beta}
 	f_1  \left(x/{\xi_\perp}, \, {t}/{\xi_\parallel}\right),
\end{equation}
where the cluster density is defined to be the coarse-grained average
density of active points.
The factor $\Delta^{\beta_1}$ comes from the probability 
that an infinite cluster can be grown from the seed, whereas $\Delta^\beta$ is 
the probability that the point (${\bf x}$, $t$) belongs to this infinite
cluster (see also \cite{grassberger-3d}). 
The shape of the cluster is governed by the scaling function $f_1$.
In (\ref{ansatz_rho_wall}) we have assumed that the density is measured at a 
finite angle $\vartheta$ away from the wall
(where $\sin \vartheta = x_\perp/x$),
and suppressed the $\vartheta$-dependence of $f_1$. 
In contrast, if the density is measured along the wall, $\vartheta = 0$, 
then the appropriate ansatz reads
\begin{equation}
	\label{ansatz_rho_wall_wall}
	\rho_{11}(x,t,\Delta) = \Delta^{2\beta_1}
  	f_{11} \left(x/{\xi_\perp}, \,
  	t/{\xi_\parallel}\right) ,
\end{equation}
as we pick up a factor $\Delta^{\beta_1}$ rather than $\Delta^\beta$
for the probability that (${\bf x}$, $t$) at the wall belongs to the
infinte cluster. In $1+1$ dimensions, $\vartheta$ has, of course, no
meaning. Instead, we have a crossover to $\rho_{11}(t,\Delta) = 
\Delta^{2\beta_1} f_{11} (t/\xi_\parallel)$ close to the wall.
We also remark that for a seed located a (finite) distance away from the 
wall, the expressions are more complicated, although the above scaling forms 
(\ref{ansatz_rho_wall}) and (\ref{ansatz_rho_wall_wall}) are still 
applicable for large times after a crossover from the bulk scaling
$\rho(x,t,\Delta) = \Delta^{2\beta} f(x/{\xi_\perp}, \, t/{\xi_\parallel})$.

By integrating the cluster density (\ref{ansatz_rho_wall}) over space and
time, we arrive at the average size of finite clusters grown from
seeds on the wall,
\begin{equation}
	\label{size_wall}
	\langle s \rangle \sim |\Delta|^{-\gamma_1} ,
\end{equation}
such that
\begin{equation}
	\nu_{\parallel}+d\nu_{\perp}=\beta_1+\beta+\gamma_{1} .
	\label{surf_hyperscaling}
\end{equation}
Hence, the surface exponent $\gamma_1$ is related to the previously
defined exponents via a scaling law that naturally generalises the usual
$d+1$ dimensional hyperscaling relation
\begin{equation}
\nu_{\parallel} +d\nu_{\perp} = 2\beta+\gamma .
\label{bulk_hyperscaling}
\end{equation}
It was noted by the authors of \cite{essam-etal:1996} that 
relation (\ref{bulk_hyperscaling}) is not fulfilled when the exponents 
for the wall geometry are substituted. Within the context of surface 
critical phenomena this ``failing'' of hyperscaling is perfectly natural, 
since (\ref{bulk_hyperscaling}) is only a relation for (the unaltered) 
bulk exponents. The results of previous numerical simulations with a wall
\cite{essam-etal:1996,lauritsen-etal} are in fact in very good agreement 
with the modified hyperscaling relation (\ref{surf_hyperscaling}).
We also note in passing that another generalisation of
hyperscaling has recently been proposed, although in the rather different
context of a (bulk) model with multiple absorbing states \cite{mendes},
where the exponent $\beta'$ in the survival probability depends
continuously on the density of the initial configuration.
Such a generalisation of hyperscaling might also apply to 
recent results in \cite{haye} for DP with different fractal seeds as 
initial conditions.

Besides integrating the density (\ref{ansatz_rho_wall}), we can also integrate
the density on the wall (\ref{ansatz_rho_wall_wall}) over the $d-1$
dimensional wall and time. This integration yields the average
(finite) cluster size on the wall,
\begin{equation}
	\label{size_wall_wall}
	\langle s_{\rm wall} \rangle \sim |\Delta|^{-\gamma_{1,1}} ,
\end{equation}
where
\begin{equation}
	\nu_{\parallel}+(d-1)\nu_{\perp}=2\beta_1+\gamma_{1,1} .
	\label{surf_hyperscaling_wall_wall}
\end{equation}
However, in higher dimensions ($d \approx 2$ being a marginal case)
this relation is not fulfilled as it would predict a negative
$\gamma_{1,1}$. For this case, $\gamma_{1,1} = 0$, reflecting a constant 
contribution to (\ref{size_wall_wall}), cf. the comment 
after (\ref{tau1_relation}).

The cluster density also contains information on the connectivity 
correlations, as it is proportional to the probability 
that the seed at the origin
is connected to the point (${\bf x}$, $t$). At criticality, we obtain 
from (\ref{ansatz_rho_wall}) the power-law decay
\begin{equation}
\label{surf_bulk_corr}
\rho_{1}(x,t) = x^{-(\beta_1 + \beta)/\nu_\perp} \tilde f_1 \left
(t\right/x^z), 
\end{equation}
where $z = \nu_\parallel / \nu_\perp$ is the dynamical exponent. This is
nothing but the critical surface-bulk correlation function with
pre-factor $x^{-(d+\eta_{1,0})}$, which defines the exponent $\eta_{1,0}$ 
describing the power-law decay of correlations 
between the surface and the bulk. Hence,
\begin{equation}
\beta_1 + \beta = \nu_\perp (d+\eta_{1,0}) ,
\end{equation}
which generalises the normal DP relation 
\begin{equation}
\label{anomal_bulk}
2 \beta = \nu_\perp (d+\eta) ,
\end{equation}
with $\eta$ the anomalous dimension in the bulk.
Furthermore, by identifying (\ref{ansatz_rho_wall_wall}) with the
surface-surface correlation function, it follows that
\begin{equation}
2 \beta_1 = \nu_\perp (d+\eta_{1,1}),
\end{equation}
where $\eta_{1,1}$ is the anomalous surface dimension.
As expected, $2 \eta_{1,0} = \eta_{1,1} + \eta$.

\section{Field-Theoretical Analysis}
\label{sec:ft}

We now turn to the field theoretic description of DP with a wall and its
connections with the above scaling picture. The action appropriate for DP
with a wall at $x_{\perp}=0$ is given by \cite{janssen-etal}
\begin{eqnarray}
& & S=S_{\rm bulk}+S_{\rm surface}, \label{actiontotal} \\
& & S_{\rm bulk}=\int d^dx\int dt ~ \left(\, \bar\phi\, [\, \partial_t
-D\nabla^2 -\Delta \, ]\,\phi+{1\over 2}u\, [\, \bar\phi\phi^2-\bar\phi^2\phi\, ]\,
\right), 
\label{actionbulk} \\
& & S_{\rm surface}=\int d^{d-1}x\int dt ~ \Delta_s\, \bar\phi_s\, \phi_s.
\label{actionsurface}
\end{eqnarray}
Here $S_{\rm bulk}$ is simply the action from Reggeon field theory 
\cite{cardy-sugar}, where $\phi$ is the local activity,
$\bar\phi$ is the response field, and where we have defined 
$\phi_s=\phi ({\bf x_{\parallel}},x_{\perp}=0,t)$ and 
$\bar \phi_s =\bar\phi({\bf x_{\parallel}},x_{\perp}=0,t)$.
The surface term in $S_{\rm surface}$
corresponds to the most relevant interaction consistent with the 
symmetries of the problem and which also respects the absorbing state
condition. Alternatively we can rewrite the action $S$ 
in the form of a Langevin-type equation for the local activity 
$\phi({\bf x},t)$,
\begin{eqnarray}
& & (\partial_t-D\nabla^2-\Delta)\phi({\bf x},t)+{1\over 2}u\phi({\bf x},t)^2
+\eta({\bf x},t)=0, \label{langevin} \\
& & \langle\eta({\bf x},t)\rangle=0, \qquad \langle\eta({\bf x},t)\eta({\bf
x}',t')\rangle= u\phi({\bf x},t)\delta^d({\bf x}-{\bf x}')\delta(t-t'),
\end{eqnarray}
where $\eta({\bf x},t)$ is a Gaussian noise term. The multiplicative factor 
$\phi({\bf x},t)$ in the noise correlator reflects the fact that $\phi=0$ 
is the absorbing state. The presence of the wall implies the 
boundary condition at $x_{\perp}=0$ of $\left. D\partial_{x_{\perp}}
\phi\right|_s=\Delta_s\phi_s$. 

For the systematic analysis of DP below the upper critical dimension, 
the action (\ref{actiontotal})--(\ref{actionsurface}) remains the more 
useful description. One can show that, in the limit $t\to\infty$, the system
reaches a steady-state, where the order parameter
$\langle\phi(x_{\perp},\Delta)\rangle$ develops a profile in the
direction away from the wall:
\begin{equation}
\label{n_wall}
\langle\phi(x_{\perp},\Delta)\rangle=\Delta^{\beta} \varphi 
\left(x_\perp/\xi_\perp\right),
\end{equation}
where the exponent $\beta$
describes the density in the bulk, $x_\perp \gg \xi_\perp$, for which
the scaling function $\varphi$ is constant. The angular brackets
denote averaging with respect to the action 
(\ref{actiontotal})--(\ref{actionsurface}). Close to the wall, however, 
the order parameter scales with a different exponent than $\beta$. 
This is in analogy with surface critical phenomena for equilibrium 
statistical mechanics. The new exponent is denoted by $\beta_1$, 
and for $\Delta > 0$ it governs the other limit of the scaling function 
$\varphi$, giving
\begin{equation}
\label{n_wall2}
\langle\phi(x_\perp, \Delta)\rangle \sim \Delta^{\beta_1} 
 x_\perp^{(\beta_1 - \beta)/\nu_\perp},
\ \ x_\perp \ll \xi_\perp .
\end{equation}
It is now a standard procedure to derive 
the form of the correlation functions within the field theory. These 
expressions involve the same $\beta$-exponents as defined above in equations 
(\ref{n_wall}) and (\ref{n_wall2}), and are identical
to the scaling forms derived earlier in section 2 (with the exception of
an additional inhomogeneous term which is present for the surface--surface 
correlation function \cite{diehl}). This establishes 
that the $\beta$-exponents defined in the field-theory above are indeed
the same as
the $\beta$-exponents used earlier in the scaling theory, which were 
defined in terms of a percolation 
probability, as in equation (\ref{perc_prob}). 
Note that the vanishing of (\ref{n_wall2}) in the limit 
$x_\perp \to 0$ is simply an artifact of the continuum analysis
(on a lattice the density on the wall simply scales as 
$\Delta^{\beta_1}$). 

Turning now to other aspects of the
field theory, it is also straightforward to show (to {\it all} orders 
in perturbation theory) that the correlation length exponents are 
everywhere unchanged by the wall --- as are all the exponents 
in the {\it bulk\/} (see \cite{diehl,janssen-etal}). Furthermore the 
surface exponent $\beta_1$ is the {\it only} new exponent introduced 
by the wall. The critical exponents 
can be calculated in a perturbative $\epsilon$ expansion around the upper
critical dimension $d_c=4$. Hence, quoting from 
\cite{janssen-etal}, we have (identical to the case of DP without a boundary)
\begin{equation}
\beta=1-{\epsilon\over 6}+O(\epsilon^2), \quad
\nu_{\parallel}= 1+{\epsilon\over 12}+O(\epsilon^2), \quad
\nu_{\perp}= {1\over 2}+{\epsilon\over 16}+O(\epsilon^2),
\end{equation}
where $\epsilon=4-d$. These exponents are related via hyperscaling 
(\ref{bulk_hyperscaling}) to $\gamma$ governing the divergence of the
bulk susceptibility (average cluster size) and via (\ref{anomal_bulk}) to 
$\eta$ governing the decay of connectivity correlations at criticality.
Furthermore an $\epsilon$ expansion calculation for the surface exponent
$\beta_1$ yields \cite{janssen-etal}
\begin{equation}
\label{beta_1}
\beta_1={3\over 2}-{7\epsilon\over 48}+O(\epsilon^2). \\
\end{equation} 
{}From the field theory of \cite{janssen-etal}, it is not hard to verify that
(\ref{surf_hyperscaling}) is the appropriate generalisation of 
(\ref{bulk_hyperscaling}), relating $\beta_1$ to
\begin{equation}
\gamma_{1}={1\over 2}+{7\epsilon\over 48}+O(\epsilon^2),
\end{equation}
which in terms of the field theory describes the divergence of the 
{\it surface\/} susceptibility due to the application of an infinitesimal
bulk field. 

The above results are certainly consistent with the numerical work 
of refs. \cite{essam-etal:1996,lauritsen-etal}, where $\nu_{\parallel}$, 
$\nu_{\perp}$ were measured in the presence of a wall and found to be 
unchanged from their bulk values. The behaviour of $\beta_1$ in 
(\ref{beta_1}) is also in
qualitative agreement with the available data. Numerically, however, the
value of the exponent $\tau_1$ was found to be 
extremely close to unity in $1+1$ dimensions. This contrasts with the 
above series results, which give
\begin{equation}
\tau_1 = -{1\over 2}+{11\epsilon\over 48}+ O(\epsilon^2) \label{eq:tau1}
\end{equation}
using the scaling relation (\ref{tau1_relation}).
Note that the mean-field value of $\tau_1$ appears to be negative. This
is not in fact the case --- following the discussion after 
(\ref{tau1_relation}), we have $\tau_1=0$ in high enough dimensions. 
Nevertheless from (\ref{eq:tau1}) 
we see that the puzzle of why $\tau_1$ seems to equal unity in 
$1+1$ dimensions cannot be answered by perturbative expansions about $d_c=4$.
Therefore this feature would appear to be a special property of DP with
a wall in $1+1$ dimensions.
This conclusion is certainly in agreement with the $2+1$ dimensional 
simulations of ref.\ \cite{lauritsen-etal}.

\section{Edge Analysis}

We next turn to the case of DP in an edge geometry, where the cluster is
started on an edge. It has been known for some time that
the presence of an edge introduces new exponents, independent of those 
associated with the bulk or with a surface (see \cite{Cardy} for a
discussion in the context of equilibrium critical phenomena, or 
\cite{grassberger-3d} in the context of percolation). However such
edge geometries have not yet (to our knowledge) been analysed for the case
of DP. 

Consider a system, where we allow the wall to have an edge with an angle 
$\alpha$ at $x_{\parallel}^{(1)}=x_\perp=0$. Hence, the edge can be 
viewed as the $d-2$ dimensional cross section of two $d-1$ dimensional
walls. By placing the seed next to this edge, the surface exponent 
$\beta_1$ is replaced by the edge exponent $\beta_2(\alpha)$ 
(where of course $\beta_1 = \beta_2(\pi)$). 
Following the same arguments as before, the survival probability for 
a cluster starting from the edge has the scaling form
\begin{equation}
P_2(t,\Delta) = \Delta^{\beta_2} \, \psi_2 \left({{t}/
{\xi_{\parallel}}}\right),            \label{eq:P_2(t,Delta)}
\end{equation}
where $\psi_2$ is constant for $t \gg \xi_\parallel$. In other words,
the percolation probability scales as $P_2(\Delta) \sim \Delta^{\beta_2}$.
Furthermore, we also have the new scaling ansatz for the cluster density
\begin{equation}
\label{ansatz_rho_edge}
\rho_{2}(r,t,\Delta) = 
 	\Delta^{\beta_2 + \beta} 
 	f_2 \left( {r}/{\xi_\perp}, \, {t}/{\xi_\parallel}\right) ,
\end{equation}
where $r$ is the radial coordinate in a system of spherical polar
coordinates centred on $x_{\parallel}^{(1)}=x_\perp=0$.
This ansatz generalises
(\ref{ansatz_rho_wall}), i.e.\ it applies for directions away from the edge
and the walls. By replacing $\beta$ with $\beta_1$ or $\beta_2$, we get the
corresponding results for the density along the wall or the edge,
respectively. Moreover, in analogy with (\ref{size_wall}) and 
(\ref{surf_hyperscaling}) for seeds on a wall, we obtain the average 
(finite) size $\langle s \rangle \sim |\Delta|^{-\gamma_2}$ of clusters
grown from a seed next to an edge, by integrating
(\ref{ansatz_rho_edge}) over space and time. This yields the relation
\begin{equation}
	\nu_{\parallel}+d\nu_{\perp}=\beta_2+\beta+\gamma_{2} .
	\label{edge_hyperscaling}
\end{equation}
Similarly, by integrating the corresponding wall density over the 
$d-1$ dimensional wall and time, we obtain the average size of cluster 
activity on the wall due to a seed at an edge, 
$\langle s_{\rm wall} \rangle \sim |\Delta|^{-\gamma_{2,1}}$, with
\begin{equation}
	\nu_{\parallel}+(d-1)\nu_{\perp}=\beta_2+\beta_1+\gamma_{2,1} .
	\label{wall_hyperscaling_21}
\end{equation}
Let us once more remark that scaling relations such as 
(\ref{edge_hyperscaling}) and (\ref{wall_hyperscaling_21}) are only valid 
as long as the predicted $\gamma$-exponents are non-negative. 
Our results indicate that $\gamma_2$ should be zero for small
enough angles $\alpha$ in any dimension, and the same holds for 
$\gamma_{2,1}$ also for somewhat larger angles. In principle, we can 
also define an exponent for the average cluster size at the edge by 
$\langle s_{\rm edge} \rangle \sim |\Delta|^{-\gamma_{2,2}}$, with
$\nu_{\parallel}+(d-2)\nu_{\perp}=2\beta_2+\gamma_{2,2}$. However, 
after inspecting our numerical results in the next section, we conclude 
that $\gamma_{2,2}$ should always be zero, with the possible exception 
of $\alpha$ close to $2 \pi$ in $d=2$. We also note that the wall
geometry in section 2 is a special case, such that 
$\gamma_2(\pi) = \gamma_1$ and $\gamma_{2,1}(\pi) = \gamma_{1,1}$, 
whereas $\gamma_{2,2}$ strictly refers to the edge.

As before, it is also straightforward to identify the various cluster 
densities with correlation functions between different domains $p$ and $q$.
It follows that
\begin{equation}
	\label{anomal_general}
	\beta_p + \beta_q = \nu_\perp (d + \eta_{p,q}), \quad
	2\eta_{p,q}=(\eta_{p,p}+\eta_{q,q}),
\end{equation}
with $p$, $q = 0$ (bulk), $1$ (wall) or $2$ (edge).

We now proceed 
to calculate the exponents of this geometry in mean-field theory. Much of
this calculation can be taken over directly from \cite{Cardy} where a
similar mean-field calculation was performed for the case of an Ising model
in an edge geometry. The appropriate terms in the action (\ref{actionbulk}),
(\ref{actionsurface}) yield mean-field equations with some resemblance 
to those of the Ising case. 
Nevertheless the presence of the time derivative and a different 
non-linear term leads to some important modifications.
However if, at the mean-field level, we are 
interested in calculating either 
the equal-time two-point correlation function or the susceptibility
then we can immediately take over the results from 
\cite{Cardy}: 
\begin{equation}
\eta_{\,0,0}=\eta=0 ,
\quad \eta_{\,1,1}=2 ,
\quad \eta_{\,2,2}=2\pi/\alpha ,
\end{equation}
and
\begin{equation}
\gamma_2=1-\pi/2\alpha.
	\label{eq:gamma2}
\end{equation}
Furthermore the correlation exponents $\nu_{\parallel}$ and $\nu_{\perp}$
are again everywhere unchanged by the presence of the edge, and retain 
their usual bulk values.

However, a calculation of the density exponent $\beta_2$ requires an
analysis of the non-linear term. Hence this exponent will differ from
that in the edge Ising model. In our case, using equation 
(\ref{langevin}), the mean-field local activity for
$\Delta>0$ must obey the equation
\begin{equation}
D\nabla^2\phi+\Delta\phi-(u/2)\phi^2=0, \label{mfbetaeq}
\end{equation}
with the boundary condition that $\phi\rightarrow 2\Delta/u$ as 
$r\rightarrow\infty$. The solution has the scaling form
\begin{equation}
\phi(r)=(\Delta/u)F(r/\xi_{\perp},\alpha).
\end{equation}
As $r\rightarrow 0$ the quadratic term in Eq.\ (\ref{mfbetaeq}) can be 
neglected, hence we obtain the Ising result, with $\phi$ behaving as
$r^{\eta_{\,2,2}/2}$. Using $\nu_\perp=1/2$ we then have 
$\phi\propto\Delta^{1+\pi/2\alpha}$, and hence
\begin{equation}
\beta_2=1+\pi/ 2\alpha.
	\label{eq:beta2}
\end{equation}
As a check, we note that this satisfies (\ref{edge_hyperscaling}) and 
(\ref{anomal_general}) at the upper critical dimension $d_c=4$.
Similarly, we naively obtain the mean-field values 
$\gamma_{2,2} = 2 \gamma_{2,1} = - \pi / \alpha < 0$, which means that 
$\gamma_{2,2} = \gamma_{2,1} = 0$, as discussed above.
Of course we could become more sophisticated and use a field theoretic
approach to calculate the fluctuation corrections to all these mean-field 
values in an $\epsilon$ expansion around $d_c=4$. Nevertheless below the 
upper critical dimension, we can still expect the mean-field values, and 
their dependence on the angle $\alpha$, to be qualitatively followed.

Let us also mention that these ideas can easily be generalised in $3+1$ 
dimensions and higher, where one for example can consider a cluster 
originating from a seed at the cross-section of three walls. 
The percolation probability and cluster densities will then scale with a 
new {\it corner} exponent $\beta_3$ satisfying scaling relations analogous 
to (\ref{edge_hyperscaling}) and (\ref{anomal_general}).

\section{Simulations}

In this section we report the results of
simulations of edge DP in $2+1$ dimensions
for opening angles of $\alpha = \pi/2$, $3\pi/4$, $\pi$ and $5\pi/4$.
We use $2+1$ dimensional bond directed percolation on a bcc lattice 
where $p_c = 0.287338(3)$, and with bulk exponents
$\beta=0.584(5)$, $\nu_{\parallel}=1.295(6)$, $\nu_{\perp}=0.734(5)$,
and the dynamic exponent $z=\nu_{\parallel}/\nu_{\perp}= 
1.765(3)$ \cite{grassberger-2d,grassberger-zhang}.
In the simulations we start from one seed located
on the edge (wall for $\alpha = \pi$) and grow the DP cluster.
Typically we average over 100,000 clusters in order to reduce the
error bars to a few percent.

We measure the average position of activity
\begin{equation}
	\left< r^2 \right> = \frac{1}{N(t, \Delta)}
				\int dV \,\, r^2 \rho_{2}(r, t, \Delta)
			   = t^{2/z} \, h(t\Delta^{\nu_\parallel}) , 
					\label{eq:x^2}
\end{equation}
where $r$ is the distance from the seed and the normalisation quantity 
$N(t,\Delta)$ is the mass of the cluster at time $t$. 
Thus the average position yields the dynamic exponent $z=\nu_\parallel/
\nu_{\perp}$. Our results show that $z$ retains its bulk value in 
agreement with the theoretical prediction.
Accordingly, we can use the bulk $z$ value in our further analysis
in order to obtain better estimates for the $\beta_2$-exponents.

Next, we measure the critical survival probability which has a power-law
behaviour 
\begin{equation}
	P_2(t)\sim t^{-\beta_2/\nu_{\parallel}} ,
            \label{eq:P_2(t)}
\end{equation}
obtained from Eq.\ (\ref{eq:P_2(t,Delta)}). The same power law also describes 
the number of active sites (for surviving clusters) on the edge 
at criticality as a function of time. We also measure the probability of 
having a cluster of mass $s$ which has the critical scaling behaviour
\begin{equation}
        p(s) \sim s^{-\tau_s}  , 
                                        \label{eq:p(s)}
\end{equation}
where $\tau_s = 1 + \beta_2/ (\nu_\parallel+d\nu_\perp-\beta)$ (cf. 
\cite{lauritsen-etal}). In addition, we measure the average number of 
active sites at criticality 
(averaged over all clusters) as function of time by integrating 
(\ref{ansatz_rho_edge}) over space,
\begin{equation}
        N(t) \sim t^{d/z - \beta/\nu_{\parallel} - \beta_2/\nu_{\parallel}}. 
%                                        \label{eq:N}
\end{equation}
However if we average only over clusters which survive to infinity, then
we have instead
\begin{equation}
        N_{\rm surv}(t) \sim t^{d/z - \beta/\nu_{\parallel}}.
%                                        \label{eq:N_surv}
\end{equation}

By measuring the above quantities we can extract various estimates for
the ratio $\beta_2/\nu_\parallel$ which eventually lead to the 
estimates for $\beta_2$ listed in table \ref{table-beta}.
In tables \ref{table-tau} and \ref{table-gamma}
we list our estimates for $\tau_2$ and $\gamma_2$. These quantities are
obtained by measuring the average lifetime $\left<t\right>$
and average size $\left<s\right>$ for finite clusters for different values
of $\Delta$ and then obtaining the exponents by carrying out a power-law fit.
We observe that the results in $2+1$ dimensions qualitatively
show the behaviour expected from the mean-field predictions.
With one exception,
we confirm that the scaling relation 
\begin{equation}
	\tau_2 = \nu_\parallel - \beta_2
				\label{eq:tau2}
\end{equation}
[cf.\ the analogous expression (\ref{tau1_relation}) for a wall] 
and hyperscaling (\ref{edge_hyperscaling}) are both fulfilled when error-bars 
are taken into account. This exception occurs for the smallest angle where 
the relation (\ref{eq:tau2}) is not fulfilled.
This is because as soon as $\beta_2$ becomes larger than $\nu_\parallel$ 
(as is the case for $\alpha=\pi/2$) the above relation breaks down, and 
instead the mean cluster lifetime becomes constant (i.e. $\tau_2=0$), 
cf.\ the comment after (\ref{tau1_relation}). Using our results for $\beta_2$ 
we find that in $2+1$ dimensions $\tau_2$ will reach its mean-field
value of zero for an angle in 
between $\pi/2$ and $3\pi/4$. When $\tau_2$ approaches zero the 
correction to scaling terms in the expression for $\left<t\right>$ will 
affect the scaling making it difficult to obtain precise values.

\section{Conclusions}

We have analysed the impact of a wall or edge on a directed percolation
process in terms of surface critical phenomena. The presence of an inactive
wall results in an {\it ordinary\/} phase transition between active and
inactive surface states at the bulk critical point. A description of this 
transition requires the introduction of one further independent exponent 
in addition to those present in the bulk. We have formulated a scaling 
ansatz for clusters growing near the surface, which incorporates the 
surface effects and explains how the exponent for the survival probability 
is altered. We have also used the connection between DP and Reggeon field 
theory to justify our scaling ansatzes and to examine the nature of the 
surface exponents. It turns out that the conjecture 
for $\tau_1 = 1$ in $1+1$ dimensions cannot be explained within
the $\epsilon$ expansion. We also remark that it would be possible
to examine other surface universality classes for evidence of rational
exponents, particularly at the {\it special\/} transition. This transition 
occurs when the surface bond probabilities are enhanced such that not only
the bulk, but also the surface, is at criticality. We note that 
the transition at this (multicritical) point requires the introduction of 
{\it two\/} new independent exponents. Lastly, we have for the first
time analysed edge exponents in DP for edges with variable opening
angles. We have derived the mean-field exponents and computed
numerical values from computer simulations in $2+1$ dimensions.

\section*{Acknowledgments}

We acknowledge discussions with Mogens H.\ Jensen, Maria 
Marko\v sov\'a and Kim Sneppen.
K.~B.~L. acknowledges the support from the Danish Natural Science Research
Council and the Carlsberg Foundation.

\begin{table}[htb]
\caption{Estimates for the $\beta_2$ exponents for $2+1$ dimensional edge DP 
         together with the mean-field values. The bulk and $1+1$ dimensional 
         wall estimates \protect\cite{essam-etal:1996,lauritsen-etal} are 
         listed for reference. The mean-field value 
	 $\beta_{2}^{(\mbox{\protect\scriptsize MF})}$ 
	 is obtained from Eq.\ (\protect\ref{eq:beta2}).
	 Recall that $\beta_2(\pi)=\beta_1$.}
\vspace*{0.5cm}
\begin{tabular}{|l||c|c|c|c||c|} \hline
Angle ($\alpha$)  &  \makebox[22mm]{$\pi/2$}   &  \makebox[22mm]{$3\pi/4$}  
                  &  \makebox[22mm]{$\pi$}     &  \makebox[22mm]{$5\pi/4$}  
                  &  \makebox[22mm]{bulk} \\ \hline\hline 
$\beta_{1}^{(1+1)}$  & & & $0.7338\pm0.0001$  & &  $0.2765\pm0.0001$ \\ \hline
$\beta_{2}^{(2+1)}$  & $1.6\pm0.1$    & $1.23\pm0.07$   
                     & $1.07\pm0.05$ & $0.98\pm0.05$ & $0.584\pm0.005$\\ \hline
$\beta_{2}^{(\mbox{\protect\scriptsize MF})}$ 
                      &    2    &   5/3  &   3/2   &   7/5   &   1  \\ \hline
\end{tabular}
\vspace*{0.5cm}
\label{table-beta}
\end{table}

\begin{table}[htb]
\caption{Estimates for the $\tau_2$ exponents for $2+1$ dimensional edge DP. 
         The bulk and $1+1$ dimensional wall estimates \protect\cite{
         essam-etal:1996,lauritsen-etal} are listed for reference. 
         Note that $\tau_{2}^{(\mbox{\protect\scriptsize MF})} = 0$.
	 Recall that $\tau_2(\pi)=\tau_1$.}
\vspace*{0.5cm}
\begin{tabular}{|l||c|c|c|c||c|} \hline
Angle ($\alpha$)  &  \makebox[22mm]{$\pi/2$}   &  \makebox[22mm]{$3\pi/4$}  
                  &  \makebox[22mm]{$\pi$}     &  \makebox[22mm]{$5\pi/4$}  
                  &  \makebox[22mm]{bulk} \\ \hline\hline 
$\tau_{1}^{(1+1)}$ &  &  & $1.0002\pm0.0003$ &  & $1.4573\pm0.0002$ \\ \hline 
$\tau_{2}^{(2+1)}$ & $0.1\pm0.05$   & $0.20\pm0.05$   
                   & $0.26\pm0.02$  & $0.38\pm0.04$ & $0.711\pm0.007$ \\ \hline
\end{tabular}
\vspace*{0.5cm}
\label{table-tau}
\end{table}

\begin{table}[htb]
\caption{Estimates for the $\gamma_2$ exponents for $2+1$ dimensional edge 
         DP together with the mean-field values. The bulk and $1+1$ 
         dimensional wall estimates \protect\cite{essam-etal:1996,
         lauritsen-etal} are listed for reference. The mean-field value 
         $\gamma_{2}^{(\mbox{\protect\scriptsize MF})}$ is obtained from 
         Eq.\ (\protect\ref{eq:gamma2}).
	 Recall that $\gamma_2(\pi)=\gamma_1$.}
\vspace*{0.5cm}
\begin{tabular}{|l||c|c|c|c||c|} \hline
Angle ($\alpha$)  &  \makebox[22mm]{$\pi/2$}   &  \makebox[22mm]{$3\pi/4$}  
                  &  \makebox[22mm]{$\pi$}     &  \makebox[22mm]{$5\pi/4$}  
                  &  \makebox[22mm]{bulk} \\ \hline\hline 
$\gamma_{1}^{(1+1)}$ &  &  & $1.8207\pm0.0004$ &  & $2.2777\pm0.0001$\\ \hline 
$\gamma_{2}^{(2+1)}$ & $0.7\pm0.1$   & $1.0\pm0.1$   
                     & $1.05\pm0.02$ & $1.20\pm0.05$ & $1.592\pm0.009$\\ \hline
$\gamma_{2}^{(\mbox{\protect\scriptsize MF})}$ 
                      &  0  &  1/3  &  1/2  &  3/5  &  1  \\ \hline
\end{tabular}
\vspace*{0.5cm}
\label{table-gamma}
\end{table}

\end{document}